\documentclass[twocolumn,aps,prl,showpacs]{revtex4}
\usepackage{graphicx}
\DeclareGraphicsExtensions{ps,eps,epsf}

\newcommand{ \caruo }{\mbox{Ca$_2$RuO$_4$}}

\newcommand{ \sigsig }{\mbox{$\sigma - \sigma '$}}
\newcommand{ \sigpi }{\mbox{$\sigma - \pi '$}}
\newcommand{ \ttg }{\mbox{$t_{2g}$}}
\newcommand{ \eg }{\mbox{$e_{g}$}}

\newcommand{ \ltwo }{\mbox{$\rm L_{II}$}}
\newcommand{ \lthree }{\mbox{$\rm L_{III}$}}
\begin{document}

\title{Orbital ordering transition in \caruo\ observed with resonant x-ray diffraction}
\author{I. Zegkinoglou$^1$, J. Strempfer$^1$, C.S. Nelson$^2$, J.P. Hill$^3$,
J. Chakhalian$^1$, C. Bernhard$^1$, J.C. Lang$^4$, G. Srajer$^4$, H. Fukazawa$^5$, S. Nakatsuji$^5$, Y. Maeno$^{5,6}$, and B. Keimer$^1$}
\affiliation{$^1$ Max-Planck-Institut f\"ur Festk\"orperforschung, Heisenbergstr. 1, D-70569 Stuttgart, Germany}
\affiliation{$^2$ National Synchrotron Light Source, Brookhaven National Laboratory, Upton, New York 11973-5000, USA}
\affiliation{$^3$ Department of Physics, Brookhaven National Laboratory, Upton, New York 11973-5000, USA}
\affiliation{$^4$ Advanced Photon Source, Argonne National Laboratory,
Argonne, Illinois 60439, USA}
\affiliation{$^5$ Department of Physics, Kyoto University, Kyoto 606-8502, Japan}
\affiliation{$^6$ International Innovation Center, Kyoto University, Kyoto 606-8501, Japan}

\date{\today}

\begin{abstract}
Resonant x-ray diffraction performed at the \ltwo\ and \lthree\
absorption edges of Ru has been used to investigate the magnetic
and orbital ordering in \caruo\ single crystals. A large resonant
enhancement due to electric dipole $2p\rightarrow 4d$ transitions
is observed at the wave-vector characteristic of antiferromagnetic
ordering. Besides the previously known antiferromagnetic phase
transition at $\rm T_{N}=110$ K, an additional phase transition,
between two paramagnetic phases, is observed around 260 K. Based
on the polarization and azimuthal angle dependence of the
diffraction signal, this transition can be attributed to orbital
ordering of the Ru $t_{2g}$ electrons. The propagation vector of
the orbital order is inconsistent with some theoretical
predictions for the orbital state of \caruo.

\end{abstract}

\pacs{75.25.+z, 71.27.+a, 75.30.-m, 61.10.-i}

\maketitle

The discovery of high temperature superconductivity in layered
cuprates has stimulated a great deal of interest in the electronic
properties of transition metal oxides (TMOs) in recent years.
Among these, 4d electron systems such as the two-dimensional Mott
transition system Ca$_{2-x}$Sr$_x$RuO$_4$ are particularly
interesting. The strong correlations induced by the narrow
electron bands, the active orbital degree of freedom, and the
unconventional superconductivity discovered in the $x=2$ end
member of this system \cite{Mae94,Nak00} are some of the
properties which have motivated its extensive study. Magnetic and
orbital ordering are expected to be intimately coupled in 4d
compounds, and resonant x-ray diffraction (RXD) is well suited to
elucidate the interplay between these two degrees of freedom. RXD
experiments performed at energies close to the K-absorption edges
of transition metals have been used to study orbital and magnetic
ordering in 3d TMOs \cite{Nak02,Mur98,Gre04,Stu99}. A much larger
resonant enhancement of the diffracted intensity is expected at
the L-edges of transition metals, because the partially filled
d-electron orbitals are then directly probed by electric dipole
transitions. This can enable the direct observation of orbital
ordering.
Such experiments have recently been performed on compounds with 3d
\cite{Wil03,Dhe04,Tho04} and 5d \cite{McM03} valence electrons, but no
measurements on 4d electron materials have been reported to date.

Here we report the results of a RXD study at the L-edges of Ru, a
4d transition metal, in the Mott insulator \caruo\ \cite{Nak97}.
Several controversial predictions have been made for the ordering
of the 4d \ttg\ Ru orbitals in this system
\cite{Miz01,Ani02,Hot02,Lee02,Jun03}, which provide a strong
motivation for this investigation. We observed a pronounced
resonant enhancement at the Ru \ltwo\ (2.9685 keV) and \lthree\
(2.837 keV) edges of the magnetic scattering intensity at the
wave-vector where antiferromagnetic (AF) order had been reported
by neutron powder diffraction \cite{Bra98}. Significant resonant
intensity was observed also above the N\'eel temperature, $\rm T_N
= 110$ K, vanishing at a second phase transition at a much higher
temperature, $\rm T_{OO} = 260$ K. We attribute this transition,
which
 has not
previously been observed, to orbital ordering.
The orbital order is characterized by the same propagation vector
as the low-temperature AF state. At wave-vectors corresponding to
theoretically predicted orbital ordering patterns with larger unit
cells \cite{Hot02}, no signal was observed.

The crystal structure of \caruo\ is based on $\rm RuO_2$ layers
built up of corner-sharing $\rm RuO_6$ octahedra. At $\rm T_M=357$
K, the material undergoes a first-order transition from a
high-temperature metallic to a low-temperature insulating phase
\cite{Nak97,Ale99,Fri01}. This transition is accompanied by a
structural transition from tetragonal to orthorhombic lattice
symmetry, which leads to a contraction of the RuO$_6$ octahedra
perpendicular to the $\rm RuO_2$ layers \cite{Fri01}.
The Ru spins order antiferromagnetically below $\rm T_N=110$ K
\cite{Nak97,Cao97,Bra98}.
In the following, the wave-vector components $(hkl)$ are indexed
in the orthorhombic space-group $Pbca$. The room temperature
lattice constants are: $a=5.4097(3)$ \AA, $b=5.4924(4)$ \AA, and
$c=11.9613(6)$ \AA\ \cite{Bra98}. In this notation, the AF order
is characterized by the propagation vector (100).

\begin{figure}[htb]
\includegraphics[width=7.2cm]{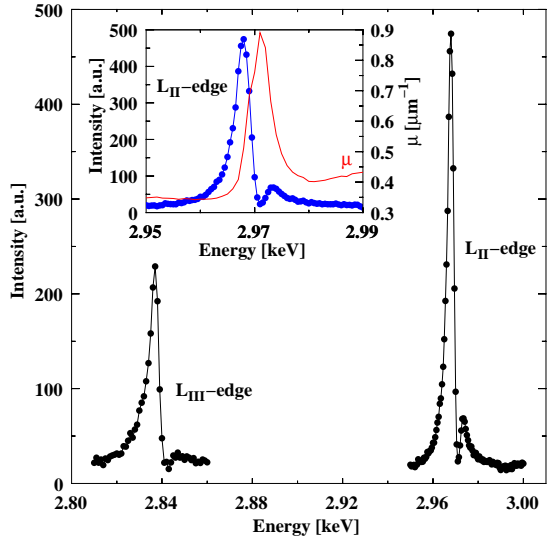}
\caption{Energy dependence of the scattered intensity of the
(100)-reflection around the \ltwo\ and \lthree\ absorption edges
at a sample temperature of 20 K ($\psi=0^{\circ}$). The energy
profiles are not corrected for absorption. The inset shows the
\ltwo\ resonance in more detail, together with the absorption
coefficient $\mu$ (right scale), calculated from fluorescence
measurements. }
  \label{fig1}
\end{figure}

The experiments were conducted at beamline 4ID-D of the Advanced
Photon Source at Argonne National Laboratory. The storage ring
chamber at this station was modified to allow measurements at
energies as low as 2.6 keV,
and the beam path was optimized to minimize the absorption of the
x-ray beam by air.
The sample was mounted in a closed-cycle cryostat capable of
reaching temperatures between 10 and 350 K, on an 8-circle
diffractometer.
A Si (111) crystal was used as polarization analyzer, providing
scattering angles of $\theta=41.8^{\circ}$ and
$\theta=44.2^{\circ}$ at the \ltwo\ and \lthree\ absorption edges,
respectively. This results in a suppression of the complementary
polarization component of the beam by 2 and 3 orders of magnitude,
respectively, for the \ltwo\ and \lthree\ edges, which is
sufficient for the separation of the two components. Two
single-crystal samples grown with the floating-zone method
\cite{Fuk01} were used for the experiments, with sizes of
approximately $100 \times 100\times 50$ $\mu m^3$ and $500 \times
400\times 50$ $\mu m^3$. The probing depth of \caruo\ in this
energy range is of the order of 2 $\mu m$.
The alignment of the samples was carried out with the $\lambda/3$
harmonic component of the incident beam using the (200) or (220)
main Bragg reflections.

\begin{figure}[htb]
\includegraphics[width=7.5cm]{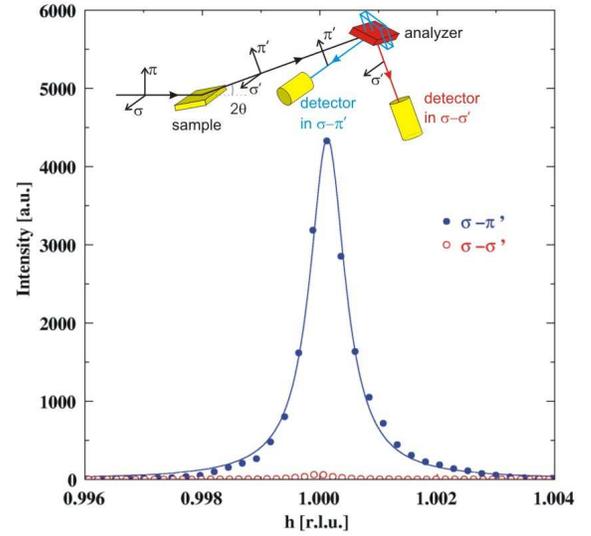}
\caption{Polarization dependence of the scattered intensity at the
(100) position at $\psi=0^{\circ}$. Data are shown for the \sigpi\
(full bullets) and \sigsig\ (empty bullets) polarization channels.
The solid line is the result of a fit to a Lorentzian profile. The inset
shows a schematic view of the experimental configuration.}
  \label{fig2}
\end{figure}

The energy dependence of the diffracted intensity at the magnetic
ordering wave-vector (100) is shown in Fig.~\ref{fig1}. The
measurements were taken at 20 K, without a polarization analyzer
and with the magnetic moment of the sample ($\vec{\mu}||\vec{b}$)
in the scattering plane. In the following, we define this
azimuthal position as $\psi=0^{\circ}$. The data shown are not
corrected for absorption. Strong resonances are observed at both
the Ru \ltwo\ and \lthree\ absorption edges. No off-resonance
scattering was observed at the (100) position within the
experimental sensitivity. The resonance enhancement at the Ru
\ltwo\ edge is thus at least a factor of 500. This enhancement
originates from electric dipole $2p\rightarrow 4d$ transitions
that probe directly the partially filled 4d orbitals responsible
for magnetism.
When the sample is rotated around the scattering vector to
$\psi=90^{\circ}$, the magnetic signal vanishes completely. The
azimuthal dependence of the scattered intensity (not shown) is
consistent with the one expected for dipole resonant magnetic
scattering \cite{Han88,Hil95}.

The inset in Fig.~\ref{fig1} shows the energy scan around the
\ltwo\ edge in more detail. In addition to the resonance peak at
the absorption edge ($\ltwo=2.9685$ keV), the data reveal a second peak, 4
eV higher in energy ($\ltwo$'$=2.9725$ keV). The origin of this peak will be
addressed below. For comparison, the solid line
in the graph shows
the energy dependence of the absorption coefficient $\mu$ around
the \ltwo\ edge, as calculated from fluorescence measurements
following Ref. \onlinecite{Bru01}.
The
inflection points of the fluorescence curve coincide with the two
peaks of the energy scan.

Reciprocal-space scans along the \textit{h}-direction conducted at
the (100) position at \ltwo\ and $\psi=0^{\circ}$ at low
temperatures are shown in Fig.~\ref{fig2}, both for the \sigpi\
and for the \sigsig\ polarization geometry. Here $\sigma$ and
$\pi$ denote the polarization components perpendicular and
parallel to the diffraction plane, respectively (inset Fig.~\ref{fig2}).
Significant
intensity is observed only in \sigpi. The weak
intensity found in \sigsig\ is due to leakage from the \sigpi\
channel ($\sim 1$\%), due to the fact that the scattering angle
of the analyzer at this energy is not
exactly 45$^{\circ}$. The absence of any \sigsig\ intensity at
$\psi=0^{\circ}$ and the absence of the reflection at
$\psi=90^{\circ}$ indicate that there is no charge scattering
contribution to the (100) intensity.

\begin{figure}[htb]
\includegraphics[width=7.5cm]{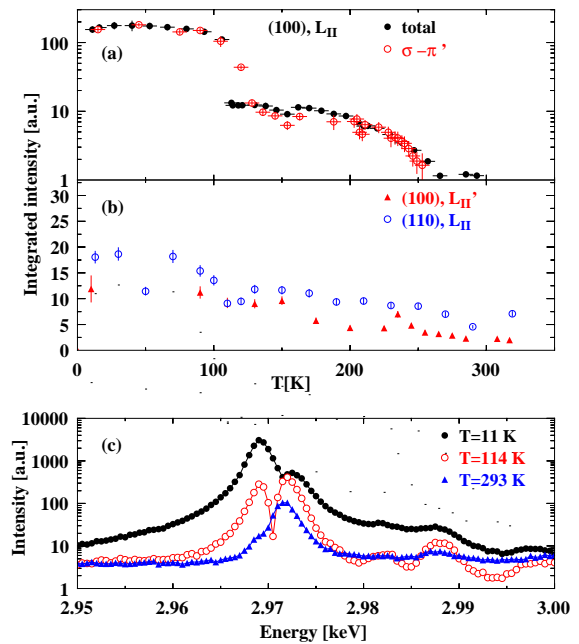}
\caption{Temperature dependence of the integrated intensity of the
(100) peak as determined from \textit{h}-scans, at (a) \ltwo, both without
analyzer and in \sigpi\ geometry (scaled), as
well as (b) at \ltwo'.
In (a), the measurements with and without polarization analyzer were taken on
two different samples. The  horizontal error bars reflect uncertainties about
the thermal coupling between sample, cold finger and temperature sensors.
In (b), because of the strong contribution
from the \ltwo\ resonance peak, the integrated intensity at \ltwo'\ could be
reliably determined from
\textit{h}-scans only above 90 K. The intensity at $\rm T=10$ K
has therefore been determined from energy scans (panel c). In (b)
the temperature dependence of the (110) peak at \ltwo\ is
also shown. (c) Variation of the intensity of the (100) peak with
energy around the \ltwo -edge for selected temperatures. All scans
were corrected for absorption.}
  \label{fig3}
\end{figure}

Fig.~\ref{fig3} shows the temperature dependence of the integrated
intensity obtained from \textit{h}-scans conducted at the (100)
position at \ltwo\, both with the analyzer in
\sigpi\ geometry and without an analyzer (Fig.~\ref{fig3}a), as
well as at \ltwo'\ without an analyzer
(Fig.~\ref{fig3}b). At \ltwo\ the previously known
magnetic transition at $\rm T_N=110$ K is clearly observed.
Remarkably, however, the intensity does not drop to zero above
$\rm T_N$. Right above $\rm T_N$, it is about a factor of 20 lower
than the intensity at 10 K. It then decreases smoothly with
increasing temperature and vanishes around 260 K in an
order-parameter-like fashion (Fig.~\ref{fig3}a; note the
logarithmic scale). This indicates a second phase transition that
has thus far not been reported. As discussed below, we attribute
the intensity above 110 K to orbital ordering of the Ru $t_{2g}$
electrons.
As is the case below $\rm T_N$, the intensity for $\rm T_N<T<260$
K is observed only in the \sigpi\ channel and vanishes when the
sample is rotated to $\rm \psi=90^{\circ}$. This means that there
is no significant charge scattering at the (100) position in this
new phase.

Energy scans conducted at temperatures below $\rm T_N$, just above
$\rm T_N$, and at $\rm T=290$ K are shown in Fig.~\ref{fig3}c. All
scans were corrected for absorption by multiplying the raw data by the
square of the energy-dependent absorption coefficient, as discussed in
Refs. \onlinecite{Ber98,Ber99}.
The line shape
above $\rm T_N$ is quite different from that below, strongly
suggesting that this new scattering is not magnetic in origin.

In order to test this, muon spin rotation ($\mu$SR)
experiments were performed on single crystals of the same
origin as the ones used in the diffraction experiments.
The measurements were carried out at the GPS beamline at the Paul
Scherrer Institute (PSI), Switzerland. No measurable static magnetic
moment was found above the N\'{e}el temperature. The detection
limit of this technique is one order of magnitude lower than the
magnitude of the magnetic moment that would be expected for the
second phase if it were magnetic, based on the intensity ratio
below and above $\rm T_N$.
The phase between $\rm T = T_N$ and $\rm T_{OO} = 260$ K is thus
paramagnetic. Based on this conclusion and on the absence of
charge scattering at (100), we attribute the 260 K transition to
the ordering of the Ru 4d \ttg\ orbitals. We note that such an
ordering, which would modify the spin Hamiltonian and hence the
paramagnetic fluctuations, may also explain anomalies in the
uniform susceptibility of \caruo\ previously reported around 260 K
\cite{Fuk01}.

\begin{figure}[htb]
\includegraphics[width=7.5cm]{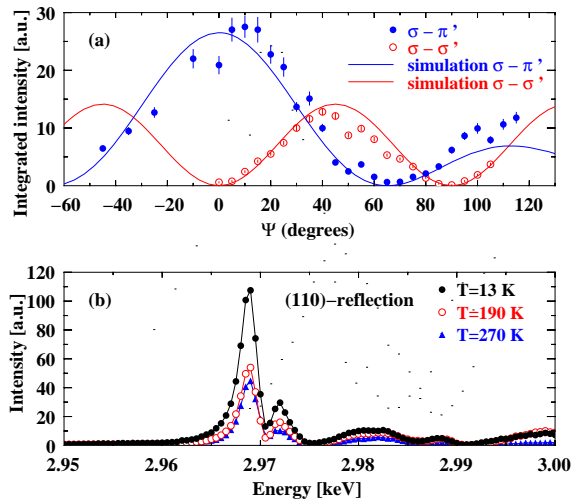}
\caption{(a) Azimuthal angle ($\psi$) dependence of the scattered
intensity of the (110) peak for both \sigpi\ (full bullets) and
\sigsig\ (empty bullets) polarization channels at T=13 K. The
solid lines are simulations of the scattered intensity induced by
the tilting of the RuO$_6$ octahedra. (b) Energy dependence of the
(110) peak intensity at $\psi=0^{\circ} $ without analyzer for different
temperatures. The data were corrected for absorption. }
  \label{fig4}
\end{figure}

The data presented thus far indicate the presence of orbital order
below  $\rm T_{OO}=260$ K characterized by the same propagation vector as
the AF
order that sets in below 110 K. Motivated by the theoretical
prediction of an orbitally ordered state with a larger unit cell
\cite{Hot02}, we also carried out searches for resonant
diffraction at reciprocal space positions consistent with the
suggested ordering pattern. However, no reflections were observed
either at (1/2 1/2 0) or at (1/2 1/2 1/2).
This scenario therefore does not appear to be viable for \caruo. While
orbital order with the observed (100) propagation vector has not
been theoretically predicted, a ``ferro-orbital" component of the
ordering pattern \cite{Lee02,Jun03,Fan04} cannot be ruled out.

Besides the new phase observed at (100) below 260 K, a resonance
peak was also found at the (110) position, which
is not allowed by the $Pbca$ space-group and is also magnetically
forbidden. The temperature dependence of the corresponding
integrated intensity is shown in Fig.~\ref{fig3}b. It is smooth,
without any phase transitions up to at least 320 K. Polarization
analysis of the diffracted intensity shows both \sigpi\ and
\sigsig\ contributions. The azimuthal dependence of both
components at \ltwo\ is shown in Fig.~\ref{fig4}a, while
Fig.~\ref{fig4}b shows the energy dependence of the total scattered
intensity at three characteristic temperatures. The latter is
similar to the one of the (100) reflection at low temperatures,
with two resonance peaks at \ltwo\ and \ltwo', but the
features observed above the edge are more pronounced here. Again,
no off-resonance intensity is found at (110).

The different temperature, polarization, and azimuthal angle
dependences of the intensities at (100) and (110) at \ltwo\
suggest that they have different origins. Previous work on 3d TMOs
has demonstrated that cooperative tilts of the metal oxide
octahedra can give rise to resonance effects at positions not
allowed by the space group \cite{Nak02}, and powder neutron
diffraction has shown that the RuO$_6$ octahedra in \caruo\ are
indeed tilted \cite{Bra98}. Based on the observed tilting pattern
and following Ref. \onlinecite{Nak02}, the azimuthal dependence of
both \sigpi\ and \sigsig\ components at (110) was computed (solid
curves in Fig.~\ref{fig4}a). The result is in good agreement with
the data. Further, the smooth temperature dependence of the tilt
angles determined by neutron diffraction \cite{Bra98} is in accord
with the lack of any anomalies observed in the (110) intensity up
to at least 320 K. This indicates that the (110) peak is due to
octahedral tilts. The same effect is also expected to give some
contribution to the intensity in the \sigpi\ polarization channel
at (100).
This may be the origin of the weak intensity remaining
above  $\rm T_{OO}$ at this position (Fig.~\ref{fig3}a).

Finally, we discuss the origin of the resonance peak observed at
\ltwo', 4 eV above the \ltwo\ edge. As indicated by the
absorption-corrected energy scans at (100) displayed in
Fig.~\ref{fig3}c, the intensity at this energy is
much more weakly temperature dependent than the one at \ltwo.
A detailed series of measurements shows that the integrated
intensity decreases gradually with increasing temperature and does
not show any anomalies up to 320 K (Fig.~\ref{fig3}b). This
suggests that the higher-energy resonance peak arises from
transitions into unoccupied orbitals not participating in the
orbital ordering transition. Based on band structure calculations,
according to which the crystal field splitting between the \ttg\
and \eg\ levels is about 4 eV \cite{Fan04}, these can be
identified as the Ru \eg\ orbitals.

In conclusion, a pronounced resonant enhancement of the scattered
intensity is observed in \caruo\ at the \ltwo\ and \lthree\
absorption edges of ruthenium. Resonant diffraction measurements
at the \ltwo\ edge reveal a sequence of phase transitions
including the previously known magnetic transition at $\rm
T_N=110$ K, as well as a new transition at  $\rm T_{OO}=260$ K.
The polarization
and temperature dependence of the intensity, combined with
supplementary $\mu$SR experiments, indicate that the latter
transition originates from ordering of the 4d \ttg\ orbitals. At
the orbital ordering wave-vector no charge scattering due to lattice
distortions is observed. This illustrates the power of resonant
x-ray diffraction to elucidate electronically driven orbital
ordering phenomena that are only weakly coupled to the crystal
lattice.

We acknowledge stimulating discussions with G. Khaliullin and
technical support from the staff at PSI.
Use of the Advanced Photon Source is supported by the U.S.
Department of Energy, Office of Basic Energy Sciences, under
contract W-31-109-Eng-38. Work at Brookhaven was supported by the
U.S. Department of Energy, Division of Materials Science, under
Contract No. DE-AC02-98CH10886. Work at Kyoto was supported in
part by Grants-in-Aid of Scientific Research from JSPS.

\bibliographystyle{prsty}

\end{document}